\def\ber{\begin{eqnarray}}
\def\eer{\end{eqnarray}}
\def\beq{\begin{equation}}
\def\eeq{\end{equation}}

\documentclass[amsmath,amssymb,nofootinbib,superscriptaddress]{revtex4}
\begin{document}

\title{Jumping from Metric $f(R)$ to Scalar-Tensor Theories and the relations between post-Newtonian parameters}

\author{Monica Capone}
\email{monica.capone@unito.it}
\affiliation{Dipartimento di Matematica,  Universit\`a di Torino, Via Carlo Alberto 10, 10125 - Torino, Italy}\affiliation{INFN, Sezione di Torino, Via Pietro Giuria 1, 10125 - Torino, Italy}

\author{Matteo Luca Ruggiero}
\email{matteo.ruggiero@polito.it}
  \affiliation{Dipartimento di Fisica, Politecnico di Torino, Corso Duca degli Abruzzi 24, 10129 - Torino, Italy} 
\affiliation{INFN, Sezione di Torino, Via Pietro Giuria 1, 10125 - Torino, Italy}

\date{\today}

\begin{abstract}
We review  the dynamical equivalence between $f(R)$  gravity in the metric formalism and scalar-tensor gravity, and use this equivalence to deduce the post-Newtonian parameters $\gamma$ and $\beta$ for a $f(R)$ theory, obtaining a result that is different with respect to that known in the literature. Then, we obtain explicit expressions of these paremeters in terms of the mass of the scalar field (or, differently speaking, the mass of the additional scalar degree of freedom associated to a $f(R)$ theory) which can be used to constrain $f(R)$ gravity by means of current observations.
\end{abstract}


\maketitle

\section{Introduction} \label{sec:intro}

In $f(R)$ theories of gravity the gravitational Lagrangian depends on an arbitrary analytic function, $f$, of the scalar curvature, $R$.  The field equations of these
theories (see \cite{capozfranc07}, \cite{Soti} and references therein) can be obtained in the metric formalism, by varying the action  with respect to the metric tensor, or in the
Palatini formalism, where the action is varied with respect to the
metric and to the affine connection, which are supposed to be independent from each other. In both cases the matter Lagrangian is supposed to depend only on the matter fields and the metric tensor. In addition to these two formalism, there is also the metric-affine formalism, in which one performs the Palatini-type variation supposing the matter Lagrangian to depend upon the metric tensor and the affine connection \cite{SLib}, besides the matter fields. These theories are also referred to as higher-order  theories of gravity, since the field equations are of higher order with respect to those of General Relativity (GR). 

The dynamical equivalence between $f(R)$ theories and a particular class of scalar-tensor (ST) theories, i.e. Brans-Dicke (BD) theory  is very well-known and studied, both in the case of metric formalism \cite{Tey & Tou}, \cite {MFF}, \cite{Wands} as well as in the Palatini formalism \cite{Fla}, \cite{Sotiriou}. We stress that two theories are said to be dynamically equivalent (limiting ourselves to the classical point of view) when it is possible to make their field equations (or actions) coincide by means of suitable redefinitions of the fields (gravitational and matter). In particular, the dynamical equivalence between $f(R)$ and the BD theory  (see e.g. \cite{Soti} and references therein) suggests to use the results known for the latter to directly obtain, after suitable manipulations, those corresponding to the former. This could be very fruitful especially for those results directly related to observations or experiments: for instance,  post-Newtonian  parameters (see \cite{ppn} for the original works, \cite{Willbook} for a detailed description of the framework and \cite{Will} for a recent review) can be used to constrain $f(R)$ theories. The issue of the post-Newtonian parameters for metric $f(R)$ theories has been previously dealt with in the literature, following the approach described in \cite{Cocoz}. However, the choice of the starting parameters and a too peculiar selection for the transformation between $f(R)$ and the corresponding BD theory have led to at least misleading results.

In this paper, after reviewing  the dynamical equivalence between $f(R)$ and ST theories, we focus on the post-Newtonian parameters issue. In particular, in the framework of metric $f(R)$ gravity,  we discuss how  post-Newtonian parameters are obtained, by exploiting the dynamical equivalence with ST theories, and we suggest that those obtained in \cite{Cocoz}  are not correct. Then we deduce the post-Newtonian parameters  $\gamma$ and $\beta$ for a metric $f(R)$ theory and, furthermore,  we give the expressions of these two parameters in terms of the mass of the scalar field, both for a general scalar-tensor theory and for a $f(R)$ theory.

\section{General correspondences between $f(R)$ gravity and scalar-tensor theories} \label{sec:gencorr}

We discuss the dynamical equivalence between $f(R)$ and ST theories, focusing on the simplest case of a fourth-order Lagrangian (see \cite {Wands}, \cite{cecotti}, \cite{Fla b} for the extension to theories of order higher than the fourth, i.e., of the type $f(R,\Box^{n}R)$, with $n\geq 1$).

By varying the action integral of $f(R)$ gravity\footnote{We use units such that $c=1$.}
\begin{equation}
S_{HO}=\frac{1}{16\pi G_{N}}\int d^{4}x\sqrt{-g}f(R)+S_{m},
\label{F(R) action}
\end{equation}
where 
\beq
S_{m}=\int d^{4}x\sqrt{-g} \mathcal{L}_{m} \label{eq:Smdef}
\eeq
is the matter action and $\mathcal{L}_{m}$ is the matter Lagrangian, with respect to the components of the metric tensor, the following fourth-order equations of motion are obtained \cite{Buch},
\begin{equation}
G_{\mu\nu}  =\frac{1}{f^{\prime}(R)}\left\{  \frac{1}{2}g_{\mu\nu}\left[
f(R)-Rf^{\prime}(R)\right]  +f^{\prime}(R)_{,\mu;\nu}-g_{\mu\nu}\square
f^{\prime}(R)\right\} +\frac{8\pi G_{N}}{f^{\prime}(R)}T_{\mu\nu} \label{Eq F(R)}
\end{equation}
where $G_{\mu\nu}$ represents the Einstein tensor and
\begin{equation}
T_{\mu\nu}\doteq-\frac{2}{\sqrt{-g}}\frac{\delta\mathcal{L}_{m}}{\delta
g^{\mu\nu}} \label{eq:tmunu}
\end{equation}
are the components of the matter/energy fields stress-energy tensor.

Now, let us consider the scalar-tensor action integral
\begin{equation}
S_{\phi}=\frac{1}{16\pi G_{N}}\int d^{4}x\sqrt{-g}\left[  f(\phi
)+(R-\phi)f^{\prime}(\phi)\right]  +S_{m}. \label{STT action 1}
\end{equation}
By varying it with respect to the metric tensor, we get the equations of motion
\begin{equation}
G_{\mu\nu}    =\frac{1}{f^{\prime}(\phi)}\left\{  \frac{1}{2}g_{\mu\nu
}\left[  f(\phi)-\phi f^{\prime}(\phi)\right]  +f^{\prime}(\phi)_{,\mu;\nu
}-g_{\mu\nu}\square f^{\prime}(\phi)\right\} +\frac{8\pi G_{N}}{f^{\prime}(\phi)}T_{\mu\nu}, \label{Eq STT}
\end{equation}
whereas a variation with respect to the scalar field, $\phi$, gives the equation 
\begin{equation}
\left[R-\phi\right]  f^{\prime\prime}(\phi)=0. \label{eq:eqscalarfield}
\end{equation}

As a consequence,  it is easy to realize that, for a given function $f$, provided that $f^{\prime\prime}(\phi)\neq0$, Eqs.(\ref{Eq STT}) coincide with
Eqs.(\ref{Eq F(R)}) \textit{on shell}, i.e. on the solutions of the last equation (\ref{eq:eqscalarfield}), so when
\begin{equation}
R=\phi. \label{R=phi}
\end{equation}

This correspondence on shell can be demonstrated to apply also in $N$
dimensions and with a more general higher-order Lagrangian, i.e. for
$(2n+4)^{th}-$order gravity \cite {Wands}, \cite{cecotti}, \cite{Fla b} (see also \cite{NOdint} for a representation of the higher-order Lagrangian density when also negative powers of the Dalambertian operator are present).

Now, it is interesting to point out that the action (\ref{STT action 1}) can be transformed into
a general ST action,
\begin{equation}
S_{ST}=\frac{1}{16\pi G_{N}}\int d^{4}x\sqrt{-g}\left[  F(\varphi
)R-Z(\varphi)g^{\mu\nu}\nabla_{\mu}\varphi\nabla_{\nu}\varphi-2V(\varphi
)\right]  +S_{m}, \label{STT gen act}
\end{equation}
by setting in the latter the identifications:
\begin{equation}
F(\varphi)=f^{\prime}(\phi),\quad Z(\varphi)=0,\text{\quad}2V(\varphi)=\phi
f^{\prime}(\phi)-f(\phi). \label{relations}
\end{equation}
Consequently, by  varying the
action (\ref{STT gen act}) with respect to both the components of the metric and the scalar field we obtain the following equations of motion, 
\begin{equation}
G_{\mu\nu} =\frac{1}{F(\varphi)}\left\{  Z(\varphi)\left[  \nabla_{\mu
}\varphi\nabla_{\nu}\varphi-\frac{1}{2}g_{\mu\nu}\nabla^{\sigma}\varphi
\nabla_{\sigma}\varphi\right]  +F(\varphi)_{,\mu;\nu}-g_{\mu\nu}\square
F(\varphi)-g_{\mu\nu}V(\varphi)\right\} +\frac{8\pi G_{N}}{F(\varphi)}T_{\mu\nu}, \label{STT gen eqs 1}
\end{equation}
and
\begin{equation}
\square\varphi+\frac{1}{2Z(\varphi)}\left\{  R\frac{dF}{d\varphi}
+\nabla^{\sigma}\varphi\nabla_{\sigma}\varphi\frac{dZ}{d\varphi}-2\frac
{dV}{d\varphi}\right\}  =0, \label{STT gen eqs 2}
\end{equation}
respectively.

As it is well-known \cite{Esp. Far.}, the action integral (\ref{STT gen act})
reduces to the generalized Brans-Dicke one by simply substituting 
\begin{equation}
F(\varphi)=\varphi,\quad Z(\varphi)=\frac{\omega}{\varphi}. \label{Fvarphi}
\end{equation}
So, on taking into account Eqs (\ref{relations}) and (\ref{Fvarphi}), we see that $f(R)$ theories can be suitably transformed into a BD theory with $\omega=0$.

In table  \ref{TableKey} a synopsis of the key steps for going to a generalized scalar-tensor Lagrangian to an $f(R)$ one are described.\\

\begin{table}[tbp] \centering
\begin{tabular}
[c]{|l|l|l|}\hline
\emph{Type} & \emph{Tricks} & $\mathcal{L}/\sqrt{-g}$\\\hline
&  & \\\hline
GST &  & $F(\varphi)R-Z(\varphi)g^{\mu\nu}\nabla_{\mu}\varphi\nabla_{\nu
}\varphi-2V(\varphi)$\\
GBD & $F(\varphi)\equiv\varphi,\ Z(\varphi)\equiv\omega/\varphi$ &
$\varphi R-\frac{\omega}{\varphi}g^{\mu\nu}\nabla_{\mu}\varphi
\nabla_{\nu}\varphi-2V(\varphi)$\\
O'H & $\omega=0$ & $\varphi R-2V(\varphi)$\\
ST & $\varphi\equiv f^{\prime}(\phi),\ 2V(\varphi)\equiv\phi f^{\prime}
(\phi)-f(\phi)$ & $f(\phi)+(R-\phi)f^{\prime}(\phi)$\\
HOT & $\phi\equiv R$ & $f(R)$\\
&  & \\\hline
\end{tabular}
\caption{In this table, the steps to follow in order to pass from a generalized scalar-tensor (GST) Lagrangian density, normalized to the square root of the metric determinant, to an $f(R)$ one are indicated. We have indicated, respectively, with GBD the generalized Brans-Dicke, O'H the O'Hanlon, ST the scalar-tensor and HOT the higher-order type normalized Lagrangian density.}\label{TableKey}
\end{table}

\textbf{Remark.}  According to what has been stated above, metric $f(R)$ gravity is nothing but a different representation of the Brans-Dicke theory with null BD parameter, $\omega=0$: as a consequence, metric $f(R)$ gravity has one extra degree of freedom with respect to General Relativity.  Actually, this extra degree of freedom is  dynamic, as one can easily deduce from the equation of motion for the $\varphi$ field, obtained  from Eq.(\ref{STT gen eqs 2}) with the substitutions (\ref{Fvarphi}), in which we set $\omega=0$, after the replacement of the Ricci scalar with the trace of Eq.(\ref{STT gen eqs 1}), that is
\begin{equation}
3\square\varphi+4V(\varphi)-2\varphi \frac{dV(\varphi)}{d\varphi}=8\pi G_{N} T.
\end{equation}
Of course, one should also bear in mind that the shape of the potential $V(\varphi)$ of the particular BD theory we are referring to is constrained by the scalar curvature. Precisely, it is
\begin{equation} \label{eq:VprimeR}
2V'(\varphi)=R,
\end{equation}
as one immediately deduces from the relations (\ref{R=phi}) and (\ref{Fvarphi}).

\section{Post-Newtonian parameters in metric $f(R)$ gravity}\label{sec:ppn}

We now focus on the deduction of the post-Newtonian parameters  for metric $f(R)$ gravity. 

To begin with, we consider the approach outlined in \cite{Damour}, where the post-Newtonian parameters for a general scalar-tensor theory of gravity are obtained, provided that we can disregard the potential associated to the scalar field. On doing so, we exploit the correspondence between $f(R)$ and scalar-tensor gravity discussed above by considering now what happens when we conformally transform a scalar-tensor and a higher-order gravity Lagrangian. It can be shown \cite{MFF}, \cite{Maeda} that both theories are
mapped into General Relativity plus a suitable number of scalar fields:  more precisely, one
can demonstrate that a scalar-tensor theory action is conformally equivalent to the Einstein-Hilbert action plus as many scalar fields as there are in the former action. On the other hand,  when a conformal transformation of an $f(R)$ action from the starting frame (referred to as \textquotedblleft Jordan
frame\textquotedblright) to the final one (called \textquotedblleft Einstein
frame\textquotedblright) is performed, one finds that it is equivalent to Einstein-Hilbert plus a scalar field action. 

Before going on, it is useful to point out that there is a longstanding debate on the issue of conformal transformations, between those who try to establish which frame is the physical one (see for example \cite{faraoni98}, \cite{MagSok} and references therein for a general discussion) and those who maintain that the two frames, under suitable redefinitions of the units, are physically indistinguishable (see \cite{Dicke CT}, \cite{Fara2}). A thorough discussion on this issue is beyond the scope of the present paper, however the reader can refer to the huge literature on this interesting problem.

Now, after performing the conformal transformation
\begin{equation}
\left\{
\begin{array}
[c]{c}
\tilde{g}_{\mu\nu}\equiv\Omega^{2}(\varphi)g_{\mu\nu}=F(\varphi)g_{\mu\nu},\\
\left(  \frac{d\psi}{d\varphi}\right)  ^{2}\equiv\frac{3}{4}\left(  \frac{d\ln
F(\varphi)}{d\varphi}\right)  ^{2}+\frac{Z(\varphi)}{2F(\varphi)},\\
A(\psi)\equiv\Omega^{-1}(\varphi)=F^{-1/2}(\varphi),\\
2U(\psi)\equiv V(\varphi)F^{-2}(\varphi),
\end{array}
\right.  \label{GTd}
\end{equation}
the action integral (\ref{STT gen act}) takes the form of the Einstein-Hilbert
action plus a scalar field \cite{Damour}, \cite{Esp. Far.},
\begin{equation}
S_{ConfTr}=\frac{1}{4\pi G_{N}}\int d^{4}x\sqrt{-g}\left[  \frac{\tilde{R}}
{4}-\frac{1}{2}\tilde{g}^{\mu\nu}\nabla_{\mu}\psi\nabla_{\nu}\psi
-U(\psi)\right]  +S_{m}\left[  \Psi_{m};A^{2}(\psi)\tilde{g}_{\mu\nu}\right], \label{eq:sconftr}
\end{equation}
where the matter term is now non-minimally coupled to the scalar field
$\psi$ through the conformal factor $A^{2}(\psi).$ Note that the new dynamical
degree of freedom $\psi$ corresponds to the starting dynamical degree of
freedom, i.e. to the scalar field $\varphi$.
On the other hand, when we conformally transform the action integral
(\ref{F(R) action}) using the conformal factor
\begin{equation}
\Omega^{2}(\varphi)=\frac{df}{dR}, \label{eq:omegaphi}
\end{equation}
the conformal transformation at stake turns out to be defined by the relations
\begin{equation}
\left\{
\begin{array}
[c]{c}
\tilde{g}_{\mu\nu}\equiv\Omega^{2}(\varphi)g_{\mu\nu}=f^{\prime}(R)g_{\mu\nu
},\\
\psi=\sqrt{\frac{3}{2}\frac{1}{8\pi G_{N}}}\ln\left(  f^{\prime}(R)\right)
,\\
A(\psi)\equiv\Omega^{-1}(\varphi)=\left[  f^{\prime}\left(  R\right)  \right]
^{-1/2},\\
2U(\psi)\equiv\frac{1}{8\pi G_{N}}\frac{Rf^{\prime}(R)-f(R)}{f^{\prime2}(R)},
\end{array}
\right.  \label{GT F(R)d}
\end{equation}
where the prime indicates the derivative with respect to the scalar curvature,
$R$. The scalar field, $\psi$, that
materializes in the Einstein frame corresponds to the
dynamical degree of freedom, $\varphi$, of the BD representation one can associate to $f(R)$ gravity. Eventually, by varying
such a transformed action, we are left with the subsequent set of equations
\begin{equation}
\tilde{G}_{\mu\nu}=2\partial_{\mu}\psi\partial_{\nu}\psi-\tilde{g}_{\mu\nu
}\partial^{\sigma}\psi\partial_{\sigma}\psi-2\tilde{g}_{\mu\nu}U(\psi)+8\pi
G_{N}\tilde{T}_{\mu\nu}, \label{STT conf transf eqs 1}
\end{equation}
for the components of the metric tensor and
\begin{equation}
\tilde{\square}\psi=-4\pi G_{N}~\alpha(\psi)\tilde{T}+\frac{dU(\psi
)}{d\psi} \label{STT conf transf eqs 2}
\end{equation}
for the scalar field, where
\begin{equation}
\alpha(\psi)\equiv\frac{d\ln A(\psi)}{d\psi} \label{eq:alphapsi}
\end{equation}
is a function giving the strength of the coupling between the scalar field and
the matter/energy source, while $\tilde{T}$ is the trace of the stress-energy
tensor\footnote{From the very definition of the stress-energy tensor also
valid in the Einstein frame, $\tilde{T}_{\mu\nu}\equiv2\ (\delta S_{m}
/\delta\tilde{g}^{\mu\nu})/\sqrt{-\tilde{g}}$, it is easy to deduce its
relation with its Jordan frame counterpart, $\tilde{T}_{\mu\nu}=A^{2}
(\psi)T_{\mu\nu}.$\ As a consequence, the contracted Bianchi identities give
us the tensorial relation $\tilde{\nabla}_{\mu}\tilde{T}_{\nu}^{\mu
}=\alpha(\psi)\tilde{T}\ \nabla_{\nu}\psi.$} back. It is worth stressing that
Eqs.(\ref{STT conf transf eqs 1}) and Eqs.(\ref{STT conf transf eqs 2}) are
exactly those one would obtain by transforming the Eqs.(\ref{STT gen eqs 1})
and (\ref{STT gen eqs 2}), respectively.

The procedure and results shown above are generalizable to actions of order
greater than the fourth. In this case, the Lagrangian (that is function of
$R,\ \square R,...,\ \square^{n}R$) is dynamically equivalent to a BD Lagrangian given in terms of $(n+1)$ scalar fields and with BD parameters all equal to zero. This last Lagrangian is conformally equivalent to General Relativity plus
$(n+1)$ scalar fields \footnote{To be more precise, we can in general define only one scalar field in the Einstein frame which turns out to have a
standard kinetic term, while the remaining degrees of freedom in Einstein
frame will have non-standard kinetic terms and may also be coupled via the
potential \cite{Wands}.}.

Having this in mind, we can proceed to determine the post-Newtonian parameters $\gamma$ and $\beta$ for a metric $f(R)$ gravity. Actually, according to the procedure described in \cite{Damour}, for a scalar-tensor action integral whose potential term can be neglected, these parameters turn out to have the following expressions:
\begin{equation}
\gamma-1=-2\frac{\alpha^{2}}{1+\alpha^{2}}\bigg|_{\psi_{0}}, \label{PPN par.s gamma 1}
\end{equation}
and
\begin{equation}
\beta-1=\frac{1}{2}\left[\frac{\alpha^{2}}{\left(1+\alpha^{2}\right)^{2}
}\frac{d\alpha}{d\psi}\right]_{\psi_{0}}, \label{PPN par.s beta}
\end{equation}
where $\psi_{0}$ is the asymptotic value of the field $\psi$. Taking into account the relations (\ref{GTd}), we can express them back into the Jordan frame. After straightforward algebra, they read:
\begin{align}
\gamma-1  &  =-\frac{(dF/d\varphi)^{2}}{ZF+2(dF/d\varphi)^{2}
}\bigg |_{\varphi_{0}},\label{gamma J.}\\
\beta-1  &  =\frac{1}{4}\left[\frac{F(dF/d\varphi)}{2ZF+3(dF/d\varphi)^{2}
}\frac{d\gamma}{d\varphi}\right]_{\varphi_{0}}, \label{beta J.}
\end{align}
where now we have to consider the asymptotic value of the field $\varphi$. These two relations, written for a general Brans-Dicke theory, give the well-known results
\begin{equation}
\gamma=\frac{1+\omega}{2+\omega}, \label{gamma Will}
\end{equation}
and
\begin{equation}
\beta-1=0, \label{beta Will}
\end{equation}
already obtained in \cite{Will} without passing through a conformal transformation.

In order to deduce the values of the post-Newtonian parameters $\gamma$ and $\beta$ for a metric $f(R)$ gravity, we start by pointing out what follows: in general, the  above parameters (\ref{PPN par.s gamma 1})-(\ref{beta J.})  are not suitable for our purposes. In fact, they have been obtained by requiring the potential associated to the scalar field of the theory to be absolutely negligible. This is not true a priori: in fact, the potential we are now considering is strictly connected to the choice of the particular $f(R)$, as it is
\begin{equation}
2V(\varphi)=Rf'(R)-f(R). \label{eq:vphiR}
\end{equation}

Actually, this fact has not been considered in \cite{Cocoz}, where the authors start from the correspondence between metric $f(R)$ and scalar-tensor theories, namely from the relations (\ref{PPN par.s gamma 1})-(\ref{beta J.}) above, to obtain the following expressions 
\begin{equation}\label{gammacapozziello}
\gamma_{R}-1\,=-\,\frac{f''(R)^2}{f'(R)+2f''(R)^2}\bigg |_{R_{0}}\,,
\end{equation}
and
\begin{equation}\label{betacapozziello}
\beta_{R}-1\,=\,\frac{1}{4}\left[\frac{f'(R)f''(R)}{2 f'(R)+3f''(R)^2}\frac{d\gamma_{R}}{d\phi}\right]_{R_{0},\phi_{0}}\,,
\end{equation}
being $R_{0}, \phi_{0}$ the asymptotic values of the scalar curvature and the $\phi$ field, respectively.

Apart from the improper use of the relations (\ref{PPN par.s gamma 1})-(\ref{beta J.}), the above results (\ref{gammacapozziello}) and (\ref{betacapozziello}), are not in agreement with those one would obtain by making use of the  substitutions that led us from a higher-order theory to the corresponding ST one,  which have bene discussed  above. In fact, by setting $F(\varphi)\equiv\varphi\equiv f^{\prime}(\phi)$ and $\phi\equiv
R$, one obtains the following expressions for the two considered post-Newtonian 
parameters:
\begin{equation}
\gamma-1=-\frac{[f^{\prime\prime}(R)\frac{dR}{d\varphi}]^{2}}
{Zf^{\prime}(R)+2[f^{\prime\prime}(R)\frac{dR}{d\varphi}]^{2}}\bigg |_{\varphi_{0}}, \label{eq:gammaPPN1}
\end{equation}
and
\begin{align}
\beta-1  &  = \frac{1}{4}\left[\frac{f^{\prime}(R)f^{\prime\prime}(R)\frac
{dR}{d\varphi}}{2Zf^{\prime}(R)+3[f^{\prime\prime}(R)\frac{dR}{d\varphi}]^{2}
}\frac{d\gamma}{dR}\frac{dR}{d\varphi}\right]_{\varphi_{0}} \label{eq:betaPPN1}\\
&  = \frac{1}{4}\left \{ \frac{f^{\prime}(R)f^{\prime\prime}(R)\frac{dR}{d\varphi}
}{2Zf^{\prime}(R)+3\left(f^{\prime\prime}(R)\frac{dR}{d\varphi}\right)^{2}}
\frac{Zf^{\prime\prime}(R)\left[ \left(f^{\prime\prime}(R)\right)^{2}-2f^{\prime
}(R)f^{\prime\prime\prime}(R)\right ]  }{\left [ Zf^{\prime}(R)+2\left(f^{\prime
\prime}(R)\frac{dR}{d\varphi}\right)^{2}\right]  ^{2}}\right\}_{\varphi_{0}}. \label{eq:betaPPN2}
\end{align}
These expressions for $\gamma$ and $\beta$ can be made equal to $\gamma_{R}$ and $\beta_{R}$, respectively, only for a specific and peculiar choice of the fields $\varphi$ and $Z(\varphi)$. In fact, the equivalence between the two sets of  relations, (\ref{gammacapozziello})-(\ref{betacapozziello}) and (\ref{eq:gammaPPN1})-(\ref{eq:betaPPN2}), only takes place when $Z=1$ and $dR=d\varphi$. In particular, the position $Z=1$ implies $f'(R)=\omega$ whereas $dR=d\varphi$ implies $f''(R)=1$. Even allowing $\omega$ to be a function of the curvature and not a constant parameter, putting together the two positions, leads us to the conclusion that the BD parameter and the $f(R)$ are obliged to be $\omega=R$ and $f(R)=(R^{2}/2)+R$, respectively\footnote{Of course, in this case, the two Eddington PPN parameters $\gamma$ and $\beta$, being constants, must be evaluated in correspondence with the asymptotic value $R_{9}$}.

Let us further comment on the issue of the interpretation of the post-Newtonian parameters (\ref{PPN par.s gamma 1}) and (\ref{PPN par.s beta}). Suppose to work with a class of higher-order theory such as  $f(R)=R^{1+\epsilon}$, with $\epsilon \to 0$.  What one would expect is that, as far as $\epsilon \to 0$, the post-Newtonian parameters reach their GR values (i.e. $\gamma=1, \beta=1$).  On the other side, the relations (\ref{gamma Will}) and (\ref{beta Will}), with $\omega=0$ (implying that it is also $Z(\varphi)=0$), immediately give the results $\gamma=1/2$ and $\beta=1$. The key point to solve this puzzle is the role of  the $Z$ function. In fact, we must consider in our discussion that the more we approach GR, the closer $Z$ approaches the infinity, and, consequently,  $\gamma \to 1$  as $\epsilon \to 0$. This further shows that a superficial application of the correspondence between $f(R)$ and scalar-tensor theories leads to results that are non self consistent.

In next section, we are going  to focus on the  issue of the relevance of the potential associated to the scalar field  when one wants to obtain the values of the post-Newtonian parameters $\gamma$ and $\beta$ for metric $f(R)$ theories starting from the correspondence with scalar-tensor gravity; as we are going to show, this fact can be related to the mass of the scalar field.

\section{Post-Newtonian parameters and the mass of the scalar field} \label{eq:ppnmass}

As pointed out in \cite{Soti} and \cite{Fara}, the ranges (or, equivalently, the masses) of the Brans-Dicke scalar field $\varphi$ can be completely different according to which definition of the mass is used. 
Of course, this choice turns out to be crucial if we aim at writing the post-Newtonian parameters $\gamma$ and $\beta$ in terms of the mass of the scalar field\footnote{Historically, the interest in the study of scalar fields provided with a mass such that the experimental limits on the $\omega$ parameter were bypassed, started after that (bosonic) string theory showed to have a low-energy limit corresponding to $\omega=-1$ Brans-Dicke theory \cite{Callan}}.  So, before going on, it is useful to clarify this issue, in order to specify which definition of mass we refer to in our approach.

To begin with, we notice that we can assign to a scalar field endowed with a  potential $V(\varphi)$ the mass
\begin{equation}
\mu^{2}(\varphi)\equiv \frac{d^{2}V}{d\varphi^{2}}, \label{mass1}
\end{equation}
that derives from the usual Klein-Gordon equation for a scalar field $\varphi$, that is
\begin{equation}
\Box \varphi-\frac{d^{2}V}{d\varphi^{2}}=S, \label{KG1}
\end{equation}
where $S$ represents the source term. Note that, with this definition, the mass turns out to be dimensionless, having both $V$ and $\varphi^{2}$ the dimensions of the fourth power of a mass\footnote{Besides $c=1$, we also set  $\hbar=1$.}. 
Actually, the field $\varphi$ that appears in the action integral of the general Brans-Dicke theory,
\begin{equation}
S_{BD}= \frac{1}{16\pi G_{N}}\int d^{4}x\sqrt{-g}\left[ \varphi R-\frac{\omega}{\varphi}
g^{\mu \nu}\nabla_{\mu}\varphi\nabla_{\nu}\varphi -2V(\varphi)\right]  +S_{m}, \label{BD action}
\end{equation}
satisfies a modified Klein-Gordon equation,
\begin{equation}
\Box \varphi=\frac{1}{3+2\omega}\left[8\pi G T+2\varphi V'(\varphi)-4V(\varphi)\right], \label{KG2}
\end{equation}
where $T$ is the trace of the matter/energy stress-energy tensor and the prime stands for derivative with respect to the $\varphi$ field. Eq.(\ref{KG2}) can be immediately recast into a Klein-Gordon one, by the introduction of an effective potential $V_{eff}(\varphi)$ such that
\begin{equation}
\frac{dV_{eff}(\varphi)}{d\varphi} =\frac{2}{3+2\omega}\left[\varphi V'(\varphi)-2V(\varphi)\right], \label{eq:dveff}
\end{equation}
so that Eq.(\ref{KG2}) becomes
\begin{equation}
\Box \varphi-\frac{dV_{eff}(\varphi)}{d\varphi}=\frac{8\pi G T}{3+2\omega}. \label{KG3}
\end{equation}
Then, an effective mass, $m({\varphi})$, can be introduced as follows:
\begin{equation}
m^{2}({\varphi})\doteq \frac{d^{2}V_{eff}}{d\varphi^{2}}=\frac{2}{3+2\omega}(\varphi V''-V'). \label{eff mass}
\end{equation}
We notice that, in this case, the mass has the proper dimensions, as now $V_{eff}$ has the dimensions of the sixth power of a mass. Thus, the correct use of the Klein-Gordon analogy seems to support the definition (\ref{eff mass}) rather than (\ref{mass1}). 

For the sake of completeness, we remember a third possible definition of mass, the one in the Einstein frame,
\begin{equation}
\tilde{m}^{2}(\psi)\equiv \frac{d^{2}U}{d\psi^{2}}.\label{mass 2}
\end{equation}
We also stress that, in this case, there are in principle as many definitions as the possible different choices of the conformal factor.

It is useful to point out that these three different  definitions do describe different physical situations, as it appears clear when considering the simple example of a constant potential $V=V_{0}\neq0$. In this case, it is $\mu=0=m({\varphi})$ thus giving an infinite range scalar field, but $\tilde{m}\neq0$, giving a finite range scalar field. 

That being said, we see that the range of $\varphi$ is completely determined by the equation of motion (\ref{KG3}) which, in the weak-field, slow-motion and spherically symmetric limit (suitable to describe the situation inside our low-density Solar System), becomes
\begin{equation}
\frac{1}{r^{2}} \frac{d}{dr}\left[r^{2}\frac{d\varphi (r)}{dr}\right]-\frac{dV_{eff}(\varphi)}{d\varphi}\simeq0. \label{SS}
\end{equation}
Now, the effective mass is obtained by expanding the potential term around the present value of $\varphi$, $\varphi_{0}$, as follows:
\begin{equation}
\frac{dV_{eff}(\varphi)}{d\varphi}\simeq\frac{dV_{eff}(\varphi)}{d\varphi}\bigg \arrowvert_{0}+
\frac{d^{2}V_{eff}(\varphi)}{d\varphi^{2}}\bigg \arrowvert_{0}\varphi=m^{2}(\varphi_{0})\varphi, \label{exp of V}
\end{equation}
where the last equality has been attained supposing $V_{eff}$ to have a minimum at $\varphi_{0}$. Equation (\ref{SS}) admits the usual Yukawa-like solution $\varphi(r)\propto \exp[-m(\varphi_{0})r]/r$ with the mass determined by the definition (\ref{eff mass}). It then appears clear that the right definition of the mass (and, consequently, the proper range for the scalar field)  to be singled out and consequently used in the  analysis of post-Newtonian parameters is Eq.(\ref{eff mass}) \cite{Fara}.

Eventually, note that from the same definition, on taking into account Eqs. (\ref{eq:VprimeR}),(\ref{eq:vphiR}), one gets the mass for the case of a generic $f(R)$ theory,
\begin{equation}
m^{2}_{HO}=\left[\frac{f'-Rf''}{3f''}\right], \label{m f(R)}
\end{equation}
provided that $f'' \neq 0$, where  primes mean derivative with respect to $R$. This result is in agreement with that found in \cite{Fara} as well as in other contests, like studies of stability \cite{stab}, perturbations \cite{pert} and propagator calculations for $f(R)$ \cite{prop}, etc. \cite{ODINI}.

For example, if we consider a Lagrangian that can be expressed as power series of the Ricci scalar around  the asymptotic value $R=0$, we may write, in general
\begin{equation}
f(R)= c_{0}+c_{1}R+c_{2}R^{2}+...+c_{n}R^{n}  \label{eq:pwseries1}
\end{equation}
where $c_{i}$ are constant coefficients. Then, from (\ref{m f(R)}), after evaluating around $R=0$, we obtain 
\beq
m^{2}_{HO}=\frac{c_{1}}{6c_{2}} \label{eq:mpwseries} 
\eeq
so that the mass is directly related to the parameters of the Lagrangian.\\

We are now able to write the post-Newtonian parameters in terms of the effective mass (\ref{eff mass}) of the scalar field, $\varphi$, and the shape of the potential, $V(\varphi)$. In fact, as we pointed out above, this is a crucial point that should be considered in order to translate the PPN parameters from a scalar-tensor to an $f(R)$ theory. As effectively stated in \cite{Periv} and in \cite{Olmo}, for a massive scalar field like the case we are discussing (where none of the two extreme conditions $m({\varphi}) r \ll 1$ and $m({\varphi}) r \gg 1$, with $r$ that indicates the scale of the experiment or observation testing the dynamics of the field, is in principle required), the correct expression of $\gamma$ is:
\begin{equation}
\gamma=\frac{3+2\omega-e^{-m_{0}r }}{3+2\omega+e^{-m_{0}r}}, \label{Periv}
\end{equation}
 where $m_{0}=m(\varphi_{0})$ is the background value of the mass, that is
 \begin{equation}
 m^{2}_{0}=\frac{2}{3+2\omega}(\varphi V''-V')\big \arrowvert_{0}.
 \end{equation}

\textbf{Remark.} The expression  (\ref{Periv}) should be interpreted as an \textit{effective} value of the post-Newtonian parameter $\gamma$, because of the dependence on the scale of the experiment $r$; as shown in \cite{Periv} and  \cite{Olmo}, this is due to a Yukawa-like correction to the Newtonian potential\footnote{It is interesting to point out that Yukawa-like corrections arise in $f(R)$ gravity also without making use of the analogy with scalar-tensor gravity (see e.g.  \cite{clifton,capozziello10}).}. In particular, it can be used to exploit the known bounds on  $\gamma$ to constrain the allowed values of $m_{0}$ and $\omega$,  on the experimental scale $r$ such that $m_{0}r \simeq 1$, so that these constraints are scale-dependent\footnote{We point out that our approach cannot be applied in principle to experiments that develop at very different scales, such as gravitational lensing; in these cases, a generalized parametrized post-Newtonian formalism should be applied that takes into account extra terms arising in $f(R)$ gravity (see e.g. \cite{clifton}).}. As for the $f(R)$ case, this fact implies that it is possible to constrain the parameters of the Lagrangian thanks to relations similar to Eq. (\ref{eq:mpwseries}). For the sake of completeness, we notice that when the background value of the mass of the scalar field is small, $m_{0} \ll 1/r$ (i.e., light, long range field), the value of the parameter $\gamma$, on which the observational bounds are directly applicable, is practically independent of the mass itself. In this case, the constraints on $\gamma$ can be turned into a constraints on $\omega$ and the relation (\ref{Periv}) reduces to the well-know one \cite{Will},
\begin{equation}
\gamma=\frac{1+\omega}{2+\omega}. \label{Will}
\end{equation}
On the other hand, when the mass is very large, $m_{0} \gg 1/r$ (i.e., very massive, short range field), the $\gamma$ parameter is inexorably driven to its GR value, that is one.\\

We can rewrite the $\omega$ parameter in terms the effective mass of the field, Eq.(\ref{eff mass}), 
\begin{equation}
\omega=\frac{1}{m^{2}({\varphi})}(\varphi V''-V')-\frac 3 2, \label{eq:omegaphi1}
\end{equation}
and substituting it into the relation (\ref{Periv}) gives
\begin{equation}
\gamma=1-2\frac{e^{-m_{0}r}}{2(\varphi V''-V')\big \arrowvert_{0}+e^{-m_{0}r}}.\label{eq:gammam}
\end{equation}
We remark that this equation is only valid in the intermediate range, that is when the mass of the scalar field is comparable to the range, $m_{0}\simeq 1/r$. 

As for the $\beta$ parameter, it can be shown \cite{Olmo}, \cite{Periv} that it is not affected by the presence of a nonnull mass for the scalar field, thus remaining frozen to the value it gets in the case of massless BD theories (as well as in GR), i.e., it is $\beta=1$.

At this point, we can finally work out the post-Newtonian parameters for a metric $f(R)$ theory of gravity. On taking into account Eqs. (\ref{m f(R)}) and (\ref{Periv}), on setting $\omega=0$ according to what we have seen in Section \ref{sec:gencorr}, we obtain
\begin{equation}
\gamma_{HO}=\frac{3-e^{-m^{0}_{HO}r}}{3+e^{-m^{0}_{HO}r}},
\end{equation}
and of course
\begin{equation}
\beta_{HO}=1
\end{equation}
where $m^{0}_{HO}$ is the background value of the mass associated to a $f(R)$ theory, given by (\ref{m f(R)}), in agreement with \cite{Olmo}.
For instance, on taking into account the solar system constraints on $\gamma$ from the Cassini mission \cite{cassini},
\beq
\gamma_{obs}=1+ \left(2.1 \pm 2.3 \right) \times 10^{-5} \label{eq:gammaobs}
\eeq
it is possible to constrain $m^{0}_{HO}$ . As an example, since in this case the scale is $r= 1 \ \mathrm{AU} \simeq 1.5 \times 10^{8} \ \mathrm{km}$, that is to say a mass scale $m_{AU} \simeq 10^{-27} \ \mathrm{GeV}$, we obtain $\frac{m^{0}_{HO}}{m_{AU}} \gtrsim 10$ which, in turn, can be used to constrain the parameters of the Lagrangian by means of Eq. (\ref{m f(R)}).

\section{Conclusions}\label{sec:conc}

We reviewed the issue of the dynamical equivalence between $f(R)$ and scalar-tensor theories, focusing on the metric formulation of $f(R)$ theory,  in order to point out how the field equations for scalar-tensor theories reduce to $f(R)$ field equations.
On exploiting this equivalence, we gave the expressions of the post-Newtonian parameters  $\gamma$ and $\beta$ for a general $f(R)$ theory, which are not the same as those available in the literature, also because the latter have been obtained by using relations that implicitly  require the potential of the scalar field to be negligible. Furthermore, we gave explicit expressions of these paremeters in terms of the mass of the scalar field, or differently speaking,  the mass of the additional scalar degree of freedom associated to an $f(R)$ theory. These expressions could be used to constrain $f(R)$ theories by means of the values of the post-Newtonian parameters obtained by the available observations and tests, as we have shown by means of a simple example. Eventually, we showed that if the mass of the scalar field is very light, values of these parameters that are in agreement with observations are obtained; on the other hand, if the scalar field is heavy, the scalar field is essentially suppressed,  and the predictions of the theory are undistinguishable from those of General Relativity. 

\section*{Acknowledgments} 

The authors would like to thank two
anonymous referees for their suggestions which contributed to
improve the paper. MC is supported by Regione Piemonte and Universit\`{a} degli Studi di Torino.

\end{document}